\documentstyle[12pt]{article}

\def\doit#1#2{\ifcase#1\or#2\fi}

\skewchar\fivmi='177
\skewchar\sixmi='177
\skewchar\sevmi='177
\skewchar\egtmi='177
\skewchar\ninmi='177
\skewchar\tenmi='177
\skewchar\elvmi='177
\skewchar\twlmi='177
\skewchar\frtnmi='177
\skewchar\svtnmi='177
\skewchar\twtymi='177
\def\@magscale#1{ scaled \magstep #1}

\def\framingfonts#1{
\doit{#1}{\font\twfvmi  = ammi10   \@magscale5 
\skewchar\twfvmi='177
\skewchar\fivsy='60
\skewchar\sixsy='60
\skewchar\sevsy='60
\skewchar\egtsy='60
\skewchar\ninsy='60
\skewchar\tensy='60
\skewchar\elvsy='60
\skewchar\twlsy='60
\skewchar\frtnsy='60
\skewchar\svtnsy='60
\skewchar\twtysy='60
\font\twfvsy  = amsy10   \@magscale5 
\skewchar\twfvsy='60
\font\go=font018			
\font\sc=font005			
\def\Go#1{{\hbox{\go #1}}}	
\def\Sc#1{{\hbox{\sc #1}}}	
\def\Sf#1{{\hbox{\sf #1}}}	
\font\oo=circlew10	      
\font\ooo=circle10			
\font\ro=manfnt				
\def\kcl{{\hbox{\ro 6}}}		
\def\kcr{{\hbox{\ro 7}}}		
\def\ktl{{\hbox{\ro \char'134}}}	
\def\ktr{{\hbox{\ro \char'135}}}	
\def\kbl{{\hbox{\ro \char'136}}}	
\def\kbr{{\hbox{\ro \char'137}}}	
}}

\catcode`@=11
\def\un#1{\relax\ifmmode\@@underline#1\else
	$\@@underline{\hbox{#1}}$\relax\fi}
\catcode`@=12

\let\du=\d			

\def\a{\alpha}
\def\b{\beta}

\def\d{\delta}
\def\e{\epsilon}

\def\g{\gamma}

\def\l{\lambda}
\def\m{\mu}
\def\n{\nu}

\def\r{\rho}
\def\s{\sigma}

\def\G{\Gamma}

\def\L{\Lambda}

\def\bo{{\raise.15ex\hbox{\large$\Box$}}}		
\def\pr{\prod}						
\def\TH{{\raise.2ex\hbox{$\displaystyle \bigodot$}\mskip-4.7mu \llap H \;}}
\def\face{{\raise.2ex\hbox{$\displaystyle \bigodot$}\mskip-2.2mu \llap {$\ddot
	\smile$}}}					

\def\Hat#1{\widehat{#1}}			
\def\Bar#1{\overline{#1}}			
\def\leftrightarrowfill{$\mathsurround=0pt \mathord\leftarrow \mkern-6mu
	\cleaders\hbox{$\mkern-2mu \mathord- \mkern-2mu$}\hfill
	\mkern-6mu \mathord\rightarrow$}
\def\dvec#1{\vbox{\ialign{##\crcr
	\leftrightarrowfill\crcr\noalign{\kern-1pt\nointerlineskip}
	$\hfil\displaystyle{#1}\hfil$\crcr}}}		

\def\frac#1#2{{\textstyle{#1\over\vphantom2\smash{\raise.20ex
	\hbox{$\scriptstyle{#2}$}}}}}			
\def\sfrac#1#2{{\vphantom1\smash{\lower.5ex\hbox{\small$#1$}}\over
	\vphantom1\smash{\raise.4ex\hbox{\small$#2$}}}}	
\def\bfrac#1#2{{\vphantom1\smash{\lower.5ex\hbox{$#1$}}\over
	\vphantom1\smash{\raise.3ex\hbox{$#2$}}}}	
\def\afrac#1#2{{\vphantom1\smash{\lower.5ex\hbox{$#1$}}\over#2}}    

\newskip\humongous \humongous=0pt plus 1000pt minus 1000pt
\def\caja{\mathsurround=0pt}
\def\eqalign#1{\,\vcenter{\openup2\jot \caja
	\ialign{\strut \hfil$\displaystyle{##}$&$
	\displaystyle{{}##}$\hfil\crcr#1\crcr}}\,}
\newif\ifdtup
\def\panorama{\global\dtuptrue \openup2\jot \caja
	\everycr{\noalign{\ifdtup \global\dtupfalse
	\vskip-\lineskiplimit \vskip\normallineskiplimit
	\else \penalty\interdisplaylinepenalty \fi}}}
\def\li#1{\panorama \tabskip=\humongous				
	\halign to\displaywidth{\hfil$\displaystyle{##}$
	\tabskip=0pt&$\displaystyle{{}##}$\hfil
	\tabskip=\humongous&\llap{$##$}\tabskip=0pt
	\crcr#1\crcr}}

\parskip=\medskipamount			
\thicklines			    

\def\oldheadpic{				
	\setlength{\unitlength}{.4mm}
	\thinlines
	\par
	\begin{picture}(349,16)
	\put(325,16){\line(1,0){4}}
	\put(330,16){\line(1,0){4}}
	\put(340,16){\line(1,0){4}}
	\put(335,0){\line(1,0){4}}
	\put(340,0){\line(1,0){4}}
	\put(345,0){\line(1,0){4}}
	\put(329,0){\line(0,1){16}}
	\put(330,0){\line(0,1){16}}
	\put(339,0){\line(0,1){16}}
	\put(340,0){\line(0,1){16}}
	\put(344,0){\line(0,1){16}}
	\put(345,0){\line(0,1){16}}
	\put(329,16){\oval(8,32)[bl]}
	\put(330,16){\oval(8,32)[br]}
	\put(339,0){\oval(8,32)[tl]}
	\put(345,0){\oval(8,32)[tr]}
	\end{picture}
	\par
	\thicklines
	\vskip.2in}
\def\oldtitle#1#2#3#4{\oldheadpic\begin{center}\vglue.5in{\large\bf #1}\\[.6in]
	{#2}\\[.1in] {\it Department of Physics and Astronomy}\\
	{\it University of Maryland, College Park, MD 20742}\\[.6in]
	Physics Publication \#{#3}\\ {#4}\\[1.5in] {\bf Abstract}\\[.1in]
	\end{center} \begin{quotation}}			
\def\oldTitle#1#2#3#4#5#6#7{\oldheadpic\begin{center} \vglue .4in
	{\large\bf #1}\\[.4in]
	{#2}\\[.1in] {\it Department of Physics and Astronomy}\\
	{\it University of Maryland, College Park, MD 20742}\\[.1in]
	{#3}\\[.1in] {\it {#4}}\\ {\it {#5}}\\[.4in]
	Physics Publication \#{#6}\\ {#7}\\[.5in] {\bf Abstract}\\[.1in]
	\end{center} \begin{quotation}}			
\def\border{						
	\setlength{\unitlength}{1mm}
	\newcount\xco
	\newcount\yco
	\xco=-24
	\yco=12
	\begin{picture}(140,0)
	\put(\xco,\yco){$\ktl$}
	\advance\yco by-1
	{\loop
	\put(\xco,\yco){$\kcl$}
	\advance\yco by-2
	\ifnum\yco>-240
	\repeat
	\put(\xco,\yco){$\kbl$}}
	\xco=158
	\yco=12
	\put(\xco,\yco){$\ktr$}
	\advance\yco by-1
	{\loop
	\put(\xco,\yco){$\kcr$}
	\advance\yco by-2
	\ifnum\yco>-240
	\repeat
	\put(\xco,\yco){$\kbr$}}
        \put(-20,11){\tiny University of Maryland Elementary Particle
Physics University of Maryland Elementary Particle Physics University of
Maryland Elementary Particle Physics}
	\put(-20,-241.5){\tiny University of Maryland Elementary
Particle Physics University of Maryland Elementary Particle Physics
University of Maryland Elementary Particle Physics}
	\end{picture}
	\par\vskip-8mm}
\def\bordero{						
	\setlength{\unitlength}{1mm}
	\newcount\xco
	\newcount\yco
	\xco=-24
	\yco=12
	\begin{picture}(140,0)
	\put(\xco,\yco){$\ktl$}
	\advance\yco by-1
	{\loop
	\put(\xco,\yco){$\kcl$}
	\advance\yco by-2
	\ifnum\yco>-240
	\repeat
	\put(\xco,\yco){$\kbl$}}
	\xco=158
	\yco=12
	\put(\xco,\yco){$\ktr$}
	\advance\yco by-1
	{\loop
	\put(\xco,\yco){$\kcr$}
	\advance\yco by-2
	\ifnum\yco>-240
	\repeat
	\put(\xco,\yco){$\kbr$}}
	\put(-20,12){\ooo
bacdefghidfghghdhededbihdgdfdfhhdheidhdhebaaahjhhdahbahgdedgehgfdiehhgdigicba}
	\put(-20,-241.5){\ooo
ababaighefdbfghgeahgdfgafagihdidihiidhiagfedhadbfdecdcdfagdcbhaddhbgfchbgfdacfediacbabab}
	\end{picture}
	\par\vskip-8mm}
\def\headpic{						
	\indent
	\setlength{\unitlength}{.4mm}
	\thinlines
	\par
	\begin{picture}(29,16)
	\put(165,16){\line(1,0){4}}
	\put(170,16){\line(1,0){4}}
	\put(180,16){\line(1,0){4}}
	\put(175,0){\line(1,0){4}}
	\put(180,0){\line(1,0){4}}
	\put(185,0){\line(1,0){4}}
	\put(169,0){\line(0,1){16}}
	\put(170,0){\line(0,1){16}}
	\put(179,0){\line(0,1){16}}
	\put(180,0){\line(0,1){16}}
	\put(184,0){\line(0,1){16}}
	\put(185,0){\line(0,1){16}}
	\put(169,16){\oval(8,32)[bl]}
	\put(170,16){\oval(8,32)[br]}
	\put(179,0){\oval(8,32)[tl]}
	\put(185,0){\oval(8,32)[tr]}
	\end{picture}
	\par\vskip-6.5mm
	\thicklines}
\def\title#1#2#3#4{\border\headpic {\hbox to\hsize{#4 \hfill UMDEPP #3}}\par
	\begin{center} \vglue .5in {\large\bf #1}\\[.6in]
	{#2}\\[.1in] {\it Department of Physics and Astronomy}\\
	{\it University of Maryland, College Park, MD 20742}\\[1.5in]
	{\bf Abstract}\\[.1in] \end{center} \begin{quotation}}	
\def\Title#1#2#3#4#5#6#7{\border\headpic
	{\hbox to\hsize{#7 \hfill UMDEPP #6}}\par
	\begin{center} \vglue .4in {\large\bf #1}\\[.4in]
	{#2}\\[.1in] {\it Department of Physics and Astronomy}\\
	{\it University of Maryland, College Park, MD 20742}\\[.1in]
	{#3}\\[.1in] {\it {#4}}\\ {\it {#5}}\\[.5in] {\bf Abstract}\\[.1in]
	\end{center} \begin{quotation}}			
\def\endtitle{\end{quotation}\newpage}			

\def\sect#1{\bigskip\medskip \goodbreak \noindent{\bf {#1}} \nobreak \medskip}
\def\refs{\sect{References} \footnotesize \frenchspacing \parskip=0pt}
\def\Item{\par\hang\textindent}

\def\doit#1#2{\ifcase#1\or#2\fi}
\def\[{\lfloor{\hskip 0.35pt}\!\!\!\lceil}
\def\]{\rfloor{\hskip 0.35pt}\!\!\!\rceil}

\def\Lag{{\cal L}}
\def\du#1#2{_{#1}{}^{#2}}
\def\ud#1#2{^{#1}{}_{#2}}
\def\dud#1#2#3{_{#1}{}^{#2}{}_{#3}}

\def\calA{{\cal A}}\def\calB{{\cal B}}\def\calD{{\cal D}}

\def\calR{{\cal R}}

\def\pl#1#2#3{Phys.~Lett.~{\bf {#1}B} (19{#2}) #3}
\def\np#1#2#3{Nucl.~Phys.~{\bf B{#1}} (19{#2}) #3}
\def\prl#1#2#3{Phys.~Rev.~Lett.~{\bf #1} (19{#2}) #3}
\def\pr#1#2#3{Phys.~Rev.~{\bf D{#1}} (19{#2}) #3}

\def\cmp#1#2#3{Comm.~Math.~Phys.~{\bf {#1}} (19{#2}) #3}

\def\ijmp#1#2#3{Int.~Jour.~Mod.~Phys.~{\bf {#1}} (19{#2}) #3}

\def\ibid#1#2#3{{\it ibid.}~{\bf {#1}} (19{#2}) #3}

\def\mpl#1#2#3{Mod.~Phys.~Lett.~{\bf A{#1}} (19{#2}) #3}

\def\half{{\fracm12}}

\def\frac#1#2{{\textstyle{#1\over\vphantom2\smash{\raise -.20ex
	\hbox{$\scriptstyle{#2}$}}}}}			

\def\fracm#1#2{\hbox{\large{${\frac{{#1}}{{#2}}}$}}}

\def\uln{{\underline n}}

\def\Hat#1{\widehat{#1}}

\def\scst{\scriptstyle}

\def\.{.$\,$}

\def\uln#1{\underline{#1}}

\def\ul{\underline}
\def\un{\underline}
\def\-{{\hskip 1.5pt}\hbox{-}}

\def\kd#1#2{\d\du{#1}{#2}}

\def\footnotew#1{\footnote{\hsize=6.5in {#1}}}

\def\low#1{{\raise -3pt\hbox{${\hskip 1.0pt}\!_{#1}$}}}

\begin{document}

\font\tenmib=cmmib10
\font\sevenmib=cmmib10 at 7pt 
\font\fivemib=cmmib10 at 5pt  
\font\tenbsy=cmbsy10
\font\sevenbsy=cmbsy10 at 7pt 
\font\fivebsy=cmbsy10 at 5pt  
\def\BMfont{\textfont0\tenbf \scriptfont0\sevenbf
                              \scriptscriptfont0\fivebf
            \textfont1\tenmib \scriptfont1\sevenmib
                               \scriptscriptfont1\fivemib
            \textfont2\tenbsy \scriptfont2\sevenbsy
                               \scriptscriptfont2\fivebsy}
\def\rlx{\relax\leavevmode}
\def\BM#1{\rlx\ifmmode\mathchoice
                      {\hbox{$\BMfont#1$}}
                      {\hbox{$\BMfont#1$}}
                      {\hbox{$\scriptstyle\BMfont#1$}}
                      {\hbox{$\scriptscriptstyle\BMfont#1$}}
                 \else{$\BMfont#1$}\fi}

\font\tenmib=cmmib10
\font\sevenmib=cmmib10 at 7pt 
\font\fivemib=cmmib10 at 5pt  
\font\tenbsy=cmbsy10
\font\sevenbsy=cmbsy10 at 7pt 
\font\fivebsy=cmbsy10 at 5pt  
\def\BMfont{\textfont0\tenbf \scriptfont0\sevenbf
                              \scriptscriptfont0\fivebf
            \textfont1\tenmib \scriptfont1\sevenmib
                               \scriptscriptfont1\fivemib
            \textfont2\tenbsy \scriptfont2\sevenbsy
                               \scriptscriptfont2\fivebsy}
\def\BM#1{\rlx\ifmmode\mathchoice
                      {\hbox{$\BMfont#1$}}
                      {\hbox{$\BMfont#1$}}
                      {\hbox{$\scriptstyle\BMfont#1$}}
                      {\hbox{$\scriptscriptstyle\BMfont#1$}}
                 \else{$\BMfont#1$}\fi}

\def\inbar{\vrule height1.5ex width.4pt depth0pt}
\def\sinbar{\vrule height1ex width.35pt depth0pt}
\def\ssinbar{\vrule height.7ex width.3pt depth0pt}
\font\cmss=cmss10
\font\cmsss=cmss10 at 7pt
\def\ZZ{\rlx\leavevmode
             \ifmmode\mathchoice
                    {\hbox{\cmss Z\kern-.4em Z}}
                    {\hbox{\cmss Z\kern-.4em Z}}
                    {\lower.9pt\hbox{\cmsss Z\kern-.36em Z}}
                    {\lower1.2pt\hbox{\cmsss Z\kern-.36em Z}}
               \else{\cmss Z\kern-.4em Z}\fi}
\def\Ik{\rlx{\rm I\kern-.18em k}}  
\def\IC{\rlx\leavevmode
             \ifmmode\mathchoice
                    {\hbox{\kern.33em\inbar\kern-.3em{\rm C}}}
                    {\hbox{\kern.33em\inbar\kern-.3em{\rm C}}}
                    {\hbox{\kern.28em\sinbar\kern-.25em{\rm C}}}
                    {\hbox{\kern.25em\ssinbar\kern-.22em{\rm C}}}
             \else{\hbox{\kern.3em\inbar\kern-.3em{\rm C}}}\fi}
\def\IP{\rlx{\rm I\kern-.18em P}}
\def\IR{\rlx{\rm I\kern-.18em R}}
\def\IN{\rlx{\rm I\kern-.20em N}}
\def\Ione{\rlx{\rm 1\kern-2.7pt l}}
\doit1{\def\ZZ{Z\!\!\!Z} \def\RR{I\!\!R} }


\def\unredoffs{} \def\redoffs{\voffset=-.31truein\hoffset=-.59truein}
\def\speclscape{\special{ps: landscape}}

\newbox\leftpage \newdimen\fullhsize \newdimen\hstitle \newdimen\hsbody
\tolerance=1000\hfuzz=2pt\def\fontflag{cm}
\catcode`\@=11 
\doit0
{
\def\bigans{b }
\message{ big or little (b/l)? }\read-1 to\answ
\ifx\answ\bigans\message{(This will come out unreduced.}
}
\hsbody=\hsize \hstitle=\hsize 
\doit0{
\else\message{(This will be reduced.} \let\l@r=L
\redoffs \hstitle=8truein\hsbody=4.75truein\fullhsize=10truein\hsize=\hsbody
\output={\ifnum\pageno=0 
  \shipout\vbox{\speclscape{\hsize\fullhsize\makeheadline}
    \hbox to \fullhsize{\hfill\pagebody\hfill}}\advancepageno
  \else
  \almostshipout{\leftline{\vbox{\pagebody\makefootline}}}\advancepageno
  \fi}
}
\def\almostshipout#1{\if L\l@r \count1=1 \message{[\the\count0.\the\count1]}
      \global\setbox\leftpage=#1 \global\let\l@r=R
 \else \count1=2
  \shipout\vbox{\speclscape{\hsize\fullhsize\makeheadline}
      \hbox to\fullhsize{\box\leftpage\hfil#1}}  \global\let\l@r=L\fi}
\fi
\def\nolabels{\def\wrlabeL##1{}\def\eqlabeL##1{}\def\reflabeL##1{}}
\def\writelabels{\def\wrlabeL##1{\leavevmode\vadjust{\rlap{\smash%
{\line{{\escapechar=` \hfill\rlap{\sevenrm\hskip.03in\string##1}}}}}}}%
\def\eqlabeL##1{{\escapechar-1\rlap{\sevenrm\hskip.05in\string##1}}}%
\def\reflabeL##1{\noexpand\llap{\noexpand\sevenrm\string\string\string##1}}}
\nolabels
%
\global\newcount\secno \global\secno=0
\global\newcount\meqno \global\meqno=1
\def\newsec#1{\global\advance\secno by1\message{(\the\secno. #1)}
\global\subsecno=0\eqnres@t\noindent{\bf\the\secno. #1}
\writetoca{{\secsym} {#1}}\par\nobreak\medskip\nobreak}
\def\eqnres@t{\xdef\secsym{\the\secno.}\global\meqno=1\bigbreak\bigskip}
\def\sequentialequations{\def\eqnres@t{\bigbreak}}\xdef\secsym{}
\global\newcount\subsecno \global\subsecno=0
\def\subsec#1{\global\advance\subsecno by1\message{(\secsym\the\subsecno. #1)}
\ifnum\lastpenalty>9000\else\bigbreak\fi
\noindent{\it\secsym\the\subsecno. #1}\writetoca{\string\quad
{\secsym\the\subsecno.} {#1}}\par\nobreak\medskip\nobreak}
\def\appendix#1#2{\global\meqno=1\global\subsecno=0\xdef\secsym{\hbox{#1.}}
\bigbreak\bigskip\noindent{\bf Appendix #1. #2}\message{(#1. #2)}
\writetoca{Appendix {#1.} {#2}}\par\nobreak\medskip\nobreak}
%
%
\def\eqnn#1{\xdef #1{(\secsym\the\meqno)}\writedef{#1\leftbracket#1}%
\global\advance\meqno by1\wrlabeL#1}
\def\eqna#1{\xdef #1##1{\hbox{$(\secsym\the\meqno##1)$}}
\writedef{#1\numbersign1\leftbracket#1{\numbersign1}}%
\global\advance\meqno by1\wrlabeL{#1$\{\}$}}
\def\eqn#1#2{\xdef #1{(\secsym\the\meqno)}\writedef{#1\leftbracket#1}%
\global\advance\meqno by1$$#2\eqno#1\eqlabeL#1$$}
%
\newskip\footskip\footskip14pt plus 1pt minus 1pt 
\def\footnotefont{\ninepoint}\def\f@t#1{\footnotefont #1\@foot}
\def\f@@t{\baselineskip\footskip\bgroup\footnotefont\aftergroup\@foot\let\next}
\setbox\strutbox=\hbox{\vrule height9.5pt depth4.5pt width0pt}
\global\newcount\ftno \global\ftno=0
\def\foot{\global\advance\ftno by1\footnote{$^{\the\ftno}$}}
%
\newwrite\ftfile
\def\footend{\def\foot{\global\advance\ftno by1\chardef\wfile=\ftfile
$^{\the\ftno}$\ifnum\ftno=1\immediate\openout\ftfile=foots.tmp\fi%
\immediate\write\ftfile{\noexpand\smallskip%
\noexpand\item{f\the\ftno:\ }\pctsign}\findarg}%
\def\footatend{\vfill\eject\immediate\closeout\ftfile{\parindent=20pt
\centerline{\bf Footnotes}\nobreak\bigskip\input foots.tmp }}}
\def\footatend{}
%
%
\global\newcount\refno \global\refno=1
\newwrite\rfile
\def\ref{[\the\refno]\nref}%
\def\nref#1{\xdef#1{[\the\refno]}\writedef{#1\leftbracket#1}%
\ifnum\refno=1\immediate\openout\rfile=refs.tmp\fi%
\global\advance\refno by1\chardef\wfile=\rfile\immediate%
\write\rfile{\noexpand\Item{#1}\reflabeL{#1\hskip.31in}\pctsign}\findarg%
\hskip10.0pt}%
\def\findarg#1#{\begingroup\obeylines\newlinechar=`\^^M\pass@rg}
{\obeylines\gdef\pass@rg#1{\writ@line\relax #1^^M\hbox{}^^M}%
\gdef\writ@line#1^^M{\expandafter\toks0\expandafter{\striprel@x #1}%
\edef\next{\the\toks0}\ifx\next\em@rk\let\next=\endgroup\else\ifx\next\empty%
\else\immediate\write\wfile{\the\toks0}\fi\let\next=\writ@line\fi\next\relax}}
\def\striprel@x#1{} \def\em@rk{\hbox{}}
\def\lref{\begingroup\obeylines\lr@f}
\def\lr@f#1#2{\gdef#1{\ref#1{#2}}\endgroup\unskip}
\def\semi{;\hfil\break}
\def\addref#1{\immediate\write\rfile{\noexpand\item{}#1}} 
\def\listrefs{\footatend\vfill\supereject\immediate\closeout\rfile\writestoppt
\baselineskip=14pt\centerline{{\bf References}}\bigskip{\frenchspacing%
\parindent=20pt\escapechar=` \input refs.tmp\vfill\eject}\nonfrenchspacing}
\def\startrefs#1{\immediate\openout\rfile=refs.tmp\refno=#1}
\def\xref{\expandafter\xr@f}\def\xr@f[#1]{#1}
\def\refs#1{\count255=1[\r@fs #1{\hbox{}}]}
\def\r@fs#1{\ifx\und@fined#1\message{reflabel \string#1 is undefined.}%
\nref#1{need to supply reference \string#1.}\fi%
\vphantom{\hphantom{#1}}\edef\next{#1}\ifx\next\em@rk\def\next{}%
\else\ifx\next#1\ifodd\count255\relax\xref#1\count255=0\fi%
\else#1\count255=1\fi\let\next=\r@fs\fi\next}
\def\figures{\centerline{{\bf Figure Captions}}\medskip\parindent=40pt%
\def\fig##1##2{\medskip\item{Fig.~##1.  }##2}}
%
\newwrite\ffile\global\newcount\figno \global\figno=1
\def\fig{fig.~\the\figno\nfig}
\def\nfig#1{\xdef#1{fig.~\the\figno}%
\writedef{#1\leftbracket fig.\noexpand~\the\figno}%
\ifnum\figno=1\immediate\openout\ffile=figs.tmp\fi\chardef\wfile=\ffile%
\immediate\write\ffile{\noexpand\medskip\noexpand\item{Fig.\ \the\figno. }
\reflabeL{#1\hskip.55in}\pctsign}\global\advance\figno by1\findarg}
\def\listfigs{\vfill\eject\immediate\closeout\ffile{\parindent40pt
\baselineskip14pt\centerline{{\bf Figure Captions}}\nobreak\medskip
\escapechar=` \input figs.tmp\vfill\eject}}
\def\xfig{\expandafter\xf@g}\def\xf@g fig.\penalty\@M\ {}
\def\figs#1{figs.~\f@gs #1{\hbox{}}}
\def\f@gs#1{\edef\next{#1}\ifx\next\em@rk\def\next{}\else
\ifx\next#1\xfig #1\else#1\fi\let\next=\f@gs\fi\next}
\newwrite\lfile
{\escapechar-1\xdef\pctsign{\string\%}\xdef\leftbracket{\string\{}
\xdef\rightbracket{\string\}}\xdef\numbersign{\string\#}}
\def\writedefs{\immediate\openout\lfile=labeldefs.tmp \def\writedef##1{%
\immediate\write\lfile{\string\def\string##1\rightbracket}}}
\def\writestop{\def\writestoppt{\immediate\write\lfile{\string\pageno%
\the\pageno\string\startrefs\leftbracket\the\refno\rightbracket%
\string\def\string\secsym\leftbracket\secsym\rightbracket%
\string\secno\the\secno\string\meqno\the\meqno}\immediate\closeout\lfile}}
\def\writestoppt{}\def\writedef#1{}
\def\seclab#1{\xdef #1{\the\secno}\writedef{#1\leftbracket#1}\wrlabeL{#1=#1}}
\def\subseclab#1{\xdef #1{\secsym\the\subsecno}%
\writedef{#1\leftbracket#1}\wrlabeL{#1=#1}}
\newwrite\tfile \def\writetoca#1{}
\def\leaderfill{\leaders\hbox to 1em{\hss.\hss}\hfill}
\def\writetoc{\immediate\openout\tfile=toc.tmp
   \def\writetoca##1{{\edef\next{\write\tfile{\noindent ##1
   \string\leaderfill {\noexpand\number\pageno} \par}}\next}}}
\def\listtoc{\centerline{\bf Contents}\nobreak\medskip{\baselineskip=12pt
 \parskip=0pt\catcode`\@=11 \input toc.tex \catcode`\@=12 \bigbreak\bigskip}}
\catcode`\@=12 
%

\def\uln#1{\underline{#1}}

\def\plpl{{+\!\!\!\!\!{\hskip 0.009in}{\raise -1.0pt\hbox{$_+$}}
{\hskip 0.0008in}}}
\def\mimi{{-\!\!\!\!\!{\hskip 0.009in}{\raise -1.0pt\hbox{$_-$}}
{\hskip 0.0008in}}}

\def\items#1{\\ \item{[#1]}}
\def\ul{\underline}
\def\un{\underline}
\def\HatD{\Hat D}
\def\kd#1#2{\d\du{#1}{#2}}

\def\framing#1{\doit{#1}
{\framingfonts{#1}
\border\headpic
}}

\framing{0}

{}~~~

\doit0{
{\bf PRELIMINARY VERSION \hfill \today} \vskip -0.05in
}
\vskip 0.07in

{\hbox to\hsize{December, 1994\hfill UMDEPP 95--66}}
{\hbox to\hsize{~~~~~ ~~~~~~\hfill HUPAPP 94--104}} \par


\begin{center}
\vglue 0.18in

{\large\bf Alternative $~N=(4,0)$~ Superstring and
$~\s\-$Models}$\,$\footnote{This  work is supported in part by NSF grant \#
PHY-93-41926.
} \\[.1in]

\baselineskip 10pt

\vskip 0.35in

\doit0{S.~James GATES, Jr.\footnote{E-mail: gates@umdhep.umd.edu}~ and~
Hitoshi NISHINO\footnote{Also at  Department of Physics and Astronomy, Howard
University, Washington, D.C. 20059, USA.
E-mail: nishino@umdhep.umd.edu.} \\[.25in]
}

\doit1{
Hitoshi ~N{\small ISHINO}\footnote{E-mail: nishino@umdhep.umd.edu.} \\[.25in]
{\it Department of Physics and Astronomy} \\[.015in]
{\it Howard University} \\[.015in]
{\it Washington, D.C. 20059, USA} \\[.18in]
and \\[.18in]
{\it Department of Physics} \\ [.015in]
{\it University of Maryland at College Park}\\ [.015in]
{\it College Park, MD 20742-4111, USA} \\[.18in]
}

\vskip 2.0in

{\bf Abstract} \\[.1in]
\end{center}

\begin{quotation}

{}~~~We present an alternative $~N=(4,0)$~ superstring theory,
with field content different from that of previously-known $~N=(4,0)$~
superstring theories.  This theory is presented as a non-linear $~\s\-$model on
the coset $~SU(n,1) / SU(n) \otimes U(1)$~ as the target space-time with
torsion, which is coupled to $~N=(4,0)$~ world-sheet superconformal gravity.
Our result indicates that the target space-time for $~N=4$~ superstring
theory does not necessarily have to be a hyper-K\"ahler or quaternionic
K\"ahler
manifold.

\endtitle

\oddsidemargin=0.03in
\evensidemargin=0.01in
\hsize=6.5in
\textwidth=6.5in

\centerline{\bf 1.~~Introduction}

The importance of superstring theories with extended supersymmetries manifests
itself in various contexts.  One exciting observation \ref\ov{H.~Ooguri
and C.~Vafa, \mpl{5}{90}{1389};
\np{361}{91}{469}; \ibid{367}{91}{83}.}\ref\ng{H.~Nishino and S.J.~Gates, Jr.,
\mpl{7}{92}{2543}.} is that $~N=2$~ superstring \ref\brink{L.~Brink and
J.H.~Schwarz, \pl{121}{77}{185}.} has self-dual (supersymmetric)
Yang-Mills theory or self-dual (super)gravity theories as its consistent
backgrounds \ref\siegelbackgrounds{W.~Siegel, \pr{47}{93}{2504}.} in its
critical four-dimensional target space-time, which play crucial roles as the
possible master theory for integrable systems in lower-dimensions
\ref\atiyah{{\it See e.g.}, M.F.~Atiyah and R.S.~Ward,
\cmp{55}{77}{117}.}\ref\kng{{\it See e.g.}, S.V.~Ketov, H.~Nishino and
S.J.~Gates, Jr., \np{393}{93}{149}; \\ S.J.~Gates, Jr.~and H.~Nishino,
\pl{299}{93}{255}; \\ H.~Nishino, \pl{316}{93}{298};
\pl{324}{94}{315}.}.  According to a recent analysis \ref\siegelcritical
{W.~Siegel, \prl{69}{92}{1493}.}, the critical dimension of
$~N=4$~ superstring theory \ref\ademollonfour{M.~Ademollo, L.~Brink, A.~D'Adda,
R.~D'Auria,  E.~Napolitano, S.~Sciuto, E.~Del Giudice, P.~Di Vecchia,
S.~Ferrara,  F.~Gliozzi, R.~Musto, R.~Pettorino, \pl{62}{76}{105};
\np{114}{76}{297}.} is probably\footnotew{The dimensionalities here are
counted in terms of bosonic scalar fields.} $~D_c=+4$~ instead of $~D_c=-8$~
which was the previous common understanding \ref\gsw{See, {\it e.g.},
M.~Green, J.H.~Schwarz and E.~Witten, {\it ``Superstring Theory''}
(Cambridge University Press, 1987)}.  If this is indeed the case, the $~N=4$~
superstring theory will gain more reason to be regarded as an important theory
like the $~N=2$~ superstring theory.

There has been recently \ref\gates{S.J.~Gates, Jr., \pl{338}{94}{31}.}
some indication that the $~N=4$~
superstring theory is not unique, but there are many different versions,
depending on the representation of the matter multiplet.  One of the reasons is
due to some ambiguity related to what is called mirror
transformation that replaces chiral superfields by twisted chiral superfields
\gates.

Independent of this indication, there has been some development
about new sets of matter multiplets with global $~N=8$~ supersymmetry
\ref\gr{S.J.~Gates, Jr.~and L.~Rana, Maryland preprints UMDEPP 94-194
and 95-060 (Oct.~1994).}, based
on what is called dimensional ``oxidation'' from one dimension $~(D=1)$~
to two-dimensions $~(D=2)$, which is a reversed procedure of the
traditional dimensional ``reduction''.  We can easily apply this
oxidation technique to get unknown matter multiplets also for the $~N=4$~
supersymmetry \gr.

In this paper, we present an alternative $~N=(4,0)$~\footnotew{In our paper we
present $~N=(p,q)$~ as $~p$~ supersymmetries with positive
chirality and $~q$~ supersymmetries with negative chirality.} matter
scalar multiplet which can be coupled to $~N=(4,0)$~ superconformal gravity
with a new field content obtained by the oxidation technique \gr\
different from the previously-known superstring
theories \ref\pn{M.~Pernici and P.~van Nieuwenhuizen,
\pl{169}{86}{886}.}\ref\bs{E.~Bergshoeff, and E.~Sezgin, \ijmp{1}{86}{191}.}.
In particular, we promote it to a heterotic non-linear $~\s\-$model on
a K\"ahler manifold \ref\ghr{S.J.~Gates, Jr.,
C.M.~Hull and M.~Ro\v cek, \np{248}{84}{157}.}, when
coupling to the $~N=(4,0)$~ supergravity.  Our results indicate that the
previously-known $~N=(4,0)$~ theories \pn\bs\ are not the only valid ones as
acceptable  superstring theories.

We first list up our possible matter scalar multiplets, which
are called SM-I through SM-IV with global $~N=(4,0)$~ supersymmetry obtained by
the dimensional oxidation.  We next couple the SM-I to $~N=(4,0)$~
superconformal gravity, with all the  scalars parametrizing the coordinates of
the coset $~SU(n,1) /SU(n) \otimes U(1)$.


\newpage

\centerline{\bf 2.~~Global $~N=(4,0)$~ Scalar Multiplets}

We first review the possible global $~N=(4,0)$~ scalar multiplets \gr, that are
obtained by the oxidation of $~D=1$~ theories, also for notational
clarification.  There are in total four
scalar multiplets, and we call them SM-I through SM-IV \gr.
All of these multiplets have $~4 + 4$~ on-shell degrees of freedom.
We give below their transformation rules and corresponding invariant
lagrangians:

\vskip 0.1in

\noindent (i) SM-I $~({\cal A},\calA^*,{\cal B},\calB^*,\psi^A)$:\footnotew{Our
notation in this paper is $~\{\g_m,\g_n\} = 2\eta_{m n} =
2\hbox{diag.}~(-,+),~\e\low{01} =
+ 1,~\g_3 = \g_0\g_1,~\g^0 = -i\s^2,~\g^1=\s^1$, so $~\g^m\g^n = \eta^{m n} +
\e^{m n} \g_3$.  We omit the indices $~{\scst +,~-}$~ for fermionic
chiralities to save space in this paper.}
$$\eqalign{&\d_Q {\cal A} = + 2 (\e^A \psi_A) \equiv + 2 (\e\psi) ~~, ~~~~
\d_Q
{\cal B} = + 2i (\Bar\e^A \psi_A) \equiv + 2i (\Bar\e \psi) ~~, \cr &\d_Q\psi^A
= - i \g^\m \Bar\e^A \partial_\m {\cal A} + \g^\m\e^A \partial_\m {\cal B} ~~,
\cr &\Lag_{\rm I} = + \left|\partial_\m {\cal A} \right|^2 +  \left|
\partial_\m
{\cal B}\right|^2 - 2i\left(\psi{}^A \g^\m\partial_\m \Bar\psi_A\right)
{}~~.\cr}
\eqno(2.1) $$
The indices $~{\scst A,~B,~\cdots~=~1,~2}$~ are for the $~{\bf
2}\-$representation of the intrinsic $~SU(2)$~ group of the $~N=(4,0)$~
superconformal gravity.  The ~${\cal A}$~ and $~{\cal B}\-$fields are complex
scalars, and $~\psi^A$~ are Weyl fermions with negative chirality, once
we chose the Weyl fermionic parameter $~\e^A$~ to have positive
chirality for the unidexterous $~N=(4,0)$~ supersymmetry.\footnotew{Note that
the charge conjugation matrix changes the chiralities: {\it e.g.}, $~\psi_{+A}=
\psi\ud-A C_{-+},~\e\du-A = \e^{+A} C_{+-}$.}  The superscript $~^*$~ denotes
its complex-conjugate.  In this paper, we usually omit the
fermionic chirality indices $~{\scst +,~-}$, unless needed for explaining
complex-conjugations, but it is useful to keep in mind that all
the gravitini carry the positive chirality.  Other conventions such as the
indices $~{\scst \m,~\n,~\cdots~=~0,~1}$~ for the world-sheet curved
coordinates are self-explanatory.
\vskip 0.1in

\noindent (ii) SM-II $~(\phi, \phi^I, \l_A, {\Bar\l}^A) $:
$$\eqalign{&\d_Q\phi = (\e^A\l_A) - ({\Bar\e}{}^A {\Bar\l}_A) \equiv
(\e\l) - ({\Bar\e}{\Bar\l}) ~~, \cr
&\d_Q \phi^I = - 2 (\e T^I \l) + 2 (\Bar\e T^I \,\Bar\l) ~~, \cr
&\d_Q \l_A = - i \g^\m \Bar\e\low A \partial_\m \phi - 2i \g^\m (T^I \Bar\e)_A
\partial_\m \phi^I ~~, \cr
&\Lag_{\rm II} = + \half (\partial_\m\phi)^2 + \half (\partial_\m\phi^I)^2
+ i\left(\Bar\l{}^A \g^\m\partial_\m  \l_A\right) ~~. \cr }
\eqno(2.2) $$
The indices $~{\scst I,~J,~\cdots~=~1,~2,~3}$~ are for the adjoint
representation for the $~SU(2)$, and $~T^I$'s are its generators
related to the Pauli's $~\s\-$matrices by $~T^I = - (i/2) \s^I$, so that
$~T^I T^J = -(1/4) \d^{I J} + (1/2) \e^{I J K} T^K$.  The action of
$~T^I$~ is such as $~(T^I\Bar\e)_A = (T^I)\du A B \Bar\e\low B = - (T^I)_{A
B} \Bar\e{}^B$.  The scalars $~\phi$~ and ~$\phi^I$~ are real, while the
fermions $~\l_A$~ are Weyl with negative chirality.
\vskip 0.1in

\noindent (iii) SM-III ~$(A_A, A^{*A}, \r, \Bar\r,\pi,\Bar\pi)$:
$$\eqalign{& \d_Q A_A = - (\e\low A \pi) + (\Bar\e\low A \r) ~~, \cr
& \d_Q \r = - 2i \g^\m \e^A \partial_\m A_A ~~, \cr
& \d_Q \pi = - 2i\g^\m \Bar\e^A \partial_\m A_A ~~, \cr
&\Lag_{\rm III} = + (\partial_\m A^{* A}) (\partial^\m A_A) + \fracm i2
\Bar\r
\g^\m\partial_\m \r + \fracm i 2 \left(\Bar\pi\g^\m\partial_\m \pi\right) ~~.
\cr}
\eqno(2.3) $$
The $~\r$~ and $~\pi$~ are both Weyl fermions with negative chirality,
while each component of $~A_A$~ is a complex scalar.
\vskip 0.1in

\noindent (iv) SM-IV $~(B_A, B^{*A}, \psi, \psi\du A B)$:
$$\eqalign{&\d_Q B_A = (\Bar\e_A \psi) - 2i (\Bar\e_B \psi\du A B) ~~, \cr
& \d_Q \psi = - i \g^\m \e^A \partial_\m B_A - i \g^\m \Bar\e_A \partial_\m
B^{* A} ~~, \cr
& \d_Q \psi\du A B = \g^\m \e^B \partial_\m B_A - \g^\m \Bar\e_A \partial_\m
B^{*B} - \half \d\du A B (\g^\m \e^C \partial_\m B_C - \g^\m \Bar\e{\,}^C
\partial_\m B^*_C ) ~~, \cr
& \Lag_{\rm IV} = + (\partial_\m B^{*A})(\partial^\m B_A) + \fracm i 2
\left(\psi\g^\m\partial_\m\psi\right)
+ i \left(\psi\du A B \g^\m\partial_\m \psi \du B A\right) ~~. \cr }
\eqno(2.4) $$
Each of $~B_A$~ is a complex scalar, while $~\psi$~ and $~\psi\du
A B$~ are Majorana-Weyl fermions with negative chirality.  The $~\du
A B\-$indices on $~\psi\du A B$~ denote the $~{\bf 3}\-$representation of
$~SU(2)$.

\bigskip\bigskip\bigskip

\centerline{\bf 3.~~SM-I Coupled to $~N=(4,0)$~ Superconformal Gravity}

Before considering $~\s\-$model couplings, we first fix the notations
for the $~N=(4,0)$~ superconformal gravity
with the field content $~(e\du\m m, \psi\du\m {+A}, {\Bar\psi}{}\dud\m
+ A, B\du \m I)$, which is compatible with the field representations of the
scalar multiplets SM-I through IV.
The transformation rule for the superconformal gravity multiplet is
$$\li{\d e\du\m m = &\, - 2i (\e\g^m \Bar\psi_\m )
  + 2i (\Bar\e\g^m \psi_\m) -\L_{\rm D} e\du\m m +\e^{m n} \L_{\rm M} e_{\m
n}~~, \cr
\d \psi\du \m A = &\, \partial _\m \e^A + \fracm 12 \omega_\m \e^A
  + B\du\m I (T^I \e)^A
  + i \g_\m \eta^A - \half \L_{\rm D} \psi\du\m A -
  \half \L_{\rm M} \psi\du\m A -  \L^I (T^I\psi_\m)^A  \cr
\equiv &\, \calD_\m\e^A
  + i \g_\m \eta^A - \half \L_{\rm D} \psi\du\m A - \half \L_{\rm M} \psi\du\m
A
  - \L^I (T^I\psi_\m)^A ~~, \cr
\d\Bar\psi_{\m A} = &\, \partial_\m \Bar\e\low A + \fracm 12
  \omega_\m\Bar\e\low A + B\du\m I (T^I \Bar\e)_A
  + i \g_\m \Bar\eta\low A - \half \L_{\rm D} \Bar\psi_{\m A}
  - \half \L_{\rm M} \Bar\psi_{\m A} - \L^I (T^I\Bar\psi_\m)_A \cr
\equiv &\, \calD_\m{\Bar\e}\low A
  + i \g_\m \Bar\eta\low A - \half \L_{\rm D} \Bar\psi_{\m A}- \half \L_{\rm M}
  \Bar\psi_{\m A} - \L^I (T^I\,\Bar\psi_\m)_A ~~,
&(3.1) \cr
\d B\du \m I = &\, +4i \left( \Bar\e \, T^I \g^\n {\cal R}_{\m\n} \right)
   - 4i \left(\e \, T^I \g^\n \Bar{\cal R}_{\m\n} \right)
   + 4\left(\Bar\psi_\n T^I \g_\m\g^\n\eta \right)
   - 4\left(\psi_\n T^I \g_\m\g^\n\Bar\eta \right)   \cr
& \, + \half \left( \d\du\m\n - e^{-1}\e\du\m\n \right) \calD_\n \L^I
   + \half \left( \d\du\m\n + e^{-1}\e\du\m\n \right) \calD_\n \L'{}^I
   ~~, \cr
\d \omega_\m = &\, - 2i (\Bar\e\g^\n \calR_{\m\n})
   + 2i (\e\g^\n {\Bar\calR}_{\m\n}) - 2(\Bar\psi_\n\g_\m\g^\n\eta)
   + 2(\psi_\n\g_\m\g^\n\Bar\eta)
   + \partial_\m \L_{\rm M} - e^{-1} \e\du\m\n \partial_\n \L_{\rm D}
{}~~. \cr } $$
The gravitino field strengths $~\calR\du{\m\n}A$~ and $~\Bar\calR_{\m\n A}$~
are
defined in terms of the covariant derivative $~\calD_\m$~ defined  in the
gravitino ~$Q\-$supertranslation.  We sometimes omit the contracted $~{\scst
A,~B,~\cdots}$~ indices, following the general rule such as ~$(\e^A \g^m
\Bar\psi_{\m A}) \equiv + \left(\e\g^m\Bar\psi_\m \right),~(\Bar\e^A \g^m
\psi_{\m A}) \equiv + \left(\Bar\e\g^m\psi_\m \right)$, {\it etc}.  The
parameters $~\L_{\rm D},~\L_{\rm M},~\L^I, ~\L'{}^I$~ are respectively the
dilatation, Lorentz, the positive and negative chirality parts of the
$~SU(2)$~ gauge transformations, while $~\omega_\m\equiv -e^{-1} \e^{\r\s}
e\du\m m \left[ \partial_\r e_{\s m} - 2i (\Bar\psi_\r\g_m\psi_\s )
\right]$~ is the Lorentz connection.

As is usual with heterotic conformal supergravity \bs, the
$~Q\-$supertranslation for the gravitino does not have any matter-dependent
terms.  This is due to the fact that the heterotic supergravity multiplet is
already off-shell with enough degrees of freedom.

We now turn to the construction of $~\s\-$model couplings of SM-I
through SM-IV to the $~N=(4,0)$~ conformal supergravity.  We see that
the SM-II and SM-IV have similar field contents, in the sense
that the representations of the scalars and fermions are flipped between
these two multiplets.  When we have a scalar field in a multiplet that has
non-trivial representation under the intrinsic $~SU(2)$~ group of the
$~N=(4,0)$~ superconformal gravity, we have to consider its minimal
coupling to the $~SU(2)$~ gauge field $~B\du\m I$.
However, as a simple trial reveals, such a minimal coupling generates
a Brans-Dicke term problematic for conformal symmetry.  To be more specific,
the variation of the $~SU(2)$~ gauge field $~B\du\m I$~ in such a minimal
coupling generates a term like $~(\e\g^\n{\calR}_{\m\n})\times (\hbox{scalar})
\,\partial_\m(\hbox{scalar})$~ with the gravitino field strength
$~\calR_{\m\n}$~ under the $~Q\-$supertranslation of the $~SU(2)$~
gauge field.  This term may be cancelled by a new term such as
$~\hbox{(fermion)}\,\g^\m\g^\n \calR_{\m\n} \times\hbox{(scalar)}$~ in the
lagrangian.  However, the $~Q\-$supertranslation of the gravitino
strength of this new term in turn necessitates a Brans-Dicke term $~R\times
(\hbox{scalar})^2$~ proportional to the scalar curvature $~R$.  There is also
some indication in superspace supporting this statement \ref\gates{S.J.~Gates,
Jr., {\it private communications}.}.

The $~\s\-$models given in refs.~\pn\bs\ are based on the
hyper-K\"ahler or quaternionic K\"ahler manifold \ref\bw{J.~Bagger and
E.~Witten, \np{222}{83}{1}.}.  In the field
representation  in \bs, we notice that the bosons and fermions are unified
in the same representation with four degrees of freedom, carrying the
curved indices of the
target space-time.  This indicates that the $~\s\-$model in \bs\
corresponds to the $~\s\-$model generalization of both SM-II and SM-IV, in
such a way that the coset structure is compatible with $~N=(4,0)$~
supersymmetry.

Among our multiplets above, we see that SM-I has not been
presented before, but has scalars in a simple representation.  In particular,
the scalar fields are singlet under the $~SU(2)$, so that they have no
minimal couplings to $~B\du\m I$.  Therefore
we do not encounter the problem mentioned above, and the coupling to
$~N=(4,0)$~
superconformal gravity must be easier.  As for the remaining SM-III,  we see
that scalar fields are in non-trivial representations under the $~SU(2)$, which
will cause the problem above.  Motivated by this this observation, we present
in
this paper the couplings of the SM-I to the $~N=(4,0)$~ superconformal gravity.

As for a possible $~\s\-$model for the SM-I, we naturally expect that its
target space-time would be a hyper-K\"ahler or quaternionic K\"ahler manifold,
according to other \hbox{$~N=4$} $\s\-$models \pn\bs.  However, a simple
counting of the degrees of freedom of the scalars and fermions reveals that no
quaternionic K\"ahler manifold seems to fit these representations.
Another subtlety is the presence of two complex scalars, whose
$~Q\-$supertranslations do not seem to be related to each other in a simple
way such as complex-conjugates.  Based on this observation, we try to loosen
the geometric restriction from the quaternionic K\"ahler manifold
to more general K\"ahler manifolds with suitable representations.  The
simplest example appears to be $~SU(n,1) / SU(n) \otimes U(1)$~ of $~n$~
complex dimensions, namely $~2n$~ real dimensions.  In fact, we
can show below that this assignment indeed works.

We use the following notation for the field content of the multiplet SM-I,
parametrizing our K\"ahler $~\s\-$model coset:
$~(\phi^\a, \Bar\phi{\,}^{\Bar\a}, \varphi^a, \varphi^*_a, \psi^{+ a A},
\Bar\psi\ud+{a A})$, where the indices  $~{\scst \a,~\b,~\cdots~=~1,~2,
{}~\cdots,~n}$
\newline
${\scst (n~\in~\{1,2,\cdots\})}$~  and their complex-conjugates
$~{\scst \Bar\a,~\Bar\b,~\cdots~=~1,~2,~\cdots,~n}$~ are for the curved
complex $~n\-$dimensional coordinates in the
coset $~SU(n,1) / SU(n) \otimes U(1)$, while the indices
$~{\scst a,~b,~\cdots~=~1,~2,~\cdots,~n}$~ are for the
$~{\bf n}\-$representations of the $~SU(n)$~ isotropy group in the
coset, and $~{\scst A,~B~=~1,~2}$~ are for the $~{\bf 2}\-$representation
of the intrinsic $~SU(2)$~ of the $~N=(4,0)$~ superconformal gravity.
The $~\phi$~ and $~\varphi\-$fields respectively correspond to the previous
scalars $~\calA$~ and $~\calB$~ in (2.1).  In particular,
the $~\varphi^a\-$field is identified as a tangent vector under the holonomy
group $~SU(n)\otimes U(1)$~ in the coset.  We sometimes use
the indices $~{\scst
\un\a~\equiv ~(\a,\,\Bar\a),~\un\b~\equiv~(\b,\,\Bar\b),~\cdots}$~ collectively
both for the barred and unbarred curved indices to save space.    Since this
coset is  essentially complex, special treatment is needed for the
complex-conjugations of fields.  The indices $~{\scst  a,~b,~\cdots}$~ as well
as $~{\scst A,~B,~\cdots}$~ change their positions from subscript to
superscript or {\it vice  versa}, under the complex-conjugations.  We will
explain more about  complex-conjugations shortly.

The vielbein on the coset $~SU(n,1) / SU(n)
\otimes U(1) $~ satisfy the usual ortho-normality conditions:
$$\eqalign{& V\du a\a V\du\a b  = \d\du a b  ~~, ~~~~V\du\a a V\du a\b =
\d\du\a \b ~~, \cr
& V^{a\Bar\a} V_{\Bar\a b} = \d\du b a ~~,~~~~V_{\Bar\a a} V^{a\Bar\b} =
\d\du{\Bar\a} {\Bar\b} ~~.  \cr }
\eqno(3.2) $$
The second line is simply the complex-conjugate of the first one,
{\it i.e.}, $~V^{a\Bar\a} = (V\du a\a)^*~$ and $~V_{\Bar\a a} = ( V\du\a
a)^*$.  An important fact is that other vielbein
components such as $~V_{\a a}$~ or $~V\du a{\Bar\a}$~ simply do {\it not}
exist in our system.  This is also related to the fact that we have {\it no}
metric for the $~SU(n)$~ indices $~{\scst
a,~b,~\cdots}$.\footnotew{The only exception is the case $~n=2$, but
even then we can manage all the relevant manipulations without a metric.}

Relevantly, our coset has the affinity $~\G\du{\un\a\un
\b}{\un\g} = \left\{{\un\g}\atop{\un\a\un\b}\right\} + T\du{\un\a\un\b}{\un\g}
$, where $~T_{\un\a\un\b\un\g}$~ is a totally antisymmetric torsion tensor
whose non-vanishing components are
$$ T_{\a\b\Bar\g} = \partial_\a B_{\b\Bar\g} - \partial_\b B_{\a\Bar\g}  ~~,
\eqno(3.3) $$
together with their complex-conjugates, where $~B_{\a\Bar\b}$~ is as usual the
antisymmetric tensor in the target space-time that appears in the
Wess-Zumino-Witten term in the $~\s\-$model.  We also notice that the
superinvariance of the $~\s\-$model action will require the relationship
$$ T\low{\a\b\Bar\g} = \half \left(\partial_\b g\low{\a\Bar\g} - \partial_\a
g\low{\b\Bar\g} \right) ~~.
\eqno(3.4) $$
This feature is similar to the $~N=(2,0)$~ heterotic $~\s\-$model
coupled to superconformal gravity \ghr\ref\bns
{E.~Bergshoeff, H.~Nishino and E.~Sezgin, \pl{165}{86}{141}.}.  Other
useful relations are such as
$$\eqalign{& \G_{\a\Bar\b\Bar\g} = 0~~, ~~~~ \G_{\a\b\g} = 0 ~~, \cr
&\G_{\a\Bar\b\g} = \partial_\a g_{\g\Bar\b} - \partial_\g g_{\a\Bar\b}
{}~~, ~~~~
\G_{\a\b\Bar\g} = \partial_\b g_{\a\Bar\g} ~~, \cr}
\eqno(3.5) $$
together with their complex-conjugates.
As usual the vielbeins are covariantly constant with respect to the affinity
$~\G\du{\un\a\un\b}{\un\g}$, and the composite connections $~A\du{\ul\a a}b,
{}~A_{\ul\a}$ respectively for the $~SU(n)$~ and $~U(1)$~ holonomy groups in
our
coset.  For instance\footnotew{Similar relations hold for the
complex-conjugates, and also for {\it barred} indices
$~{\scst\Bar\a,~\Bar\b,~\cdots}$~ that we clarify next.}
$$ \calD_\a V\du\b c \equiv \partial_\a V\du\b c
-\G\du{\a\b}\g V\du\g c  - A\du{\a c} b V\du\b c + i A_\a V\du\b c \equiv 0~~.
\eqno(3.6) $$

Some technical care is needed about the complex-conjugation rules in
our system.  When taking complex-conjugates, we have to distinguish the
basic fields from those produced by the multiplications of $~C_{A B}$~
or $~C^{A B}$.  To be more specific, our basic fields are
$$\psi^{a A}~~, ~~~~\Bar\psi_{a A}  ~~, ~~~~
\psi\du\m A ~~, ~~~~\Bar\psi_{\m A}  ~~, ~~~~\e^A  ~~, ~~~~\Bar\e\low A~~,
{}~~~~\eta^A~~, ~~~~\Bar\eta\low A~~,
\eqno(3.7) $$
while non-basic fields such as $~\psi_{\m A}$~ are
regarded as the multiplications of the corresponding basic fields by the
$~C_{A B}$~ or $~C^{A B}$: ~$\psi_{\m A} \equiv \psi\du\m B C_{B A}$.
Illustrative examples are
$$\left( \psi\du\m{+ A} \right)^* = \Bar\psi\dud \m + A ~~, ~~~~\left(
\Bar\psi\dud\m + A\right)^* = \psi\du\m{+ A} ~~, ~~~~
\left(\psi\du{\m -} A \right)^* = - \Bar\psi_{\m -A}~~, ~~~~
\left(\Bar\psi _{\m- A} \right)^* = - \psi\du{\m -} A ~~, $$
$$ \left( \psi^{-a A}\right)^* = \Bar\psi{\,}\ud- {a A}  ~~, ~~~~
\left(\Bar\psi{\,}\ud-{a A} \right) ^* = \psi^{-a A} ~~, ~~~~
\left(\psi\du+ {a A} \right)^* = - \Bar\psi_{+a A} ~~, ~~~~\left(
\Bar\psi_{+ a A}\right)^* = - \psi\du + {a A} ~~, $$
$$\left(\psi\dud\m + A\right)^* = \left(\psi\du\m{+B} C_{B A} \right)^*
= (C_{B A})^* \left(\psi\du\m{+ B}\right)^* = C^{B A} \Bar\psi\dud\m + B = -
C^{A B} \Bar\psi\dud\m + B = - \Bar\psi\du\m{+ A} ~~, $$
$$\left( \Bar\psi\du\m {+ A} \right)^* = \left(C^{A
B}\Bar\psi\dud\m+B\right)^* = (\Bar\psi\dud\m + B)^*(C^{A B})^* = \psi
\du\m{+B}
C_{A B} = - \psi\du\m{+B} C_{B A} = - \psi\dud\m + A ~~,{~~~}
\hbox{(3.8)} $$
obeying also the traditional rules in ref.~\ref\ggrs{S.J.~Gates, Jr.,
M.T.~Grisaru, M.~Ro\v cek and W.~Siegel, {\it ``Superspace''},
(Benjamin/Cummings, 1983).}.  Other helpful relations are $~\left[
(\g^\m)_{+ +} \right]^* = (\g^\m)_{+ +},~ \left[ (\g^\m)_{-\,-}
\right]^* = (\g^\m)_{-\,-},~ \left[ (T^I)\du A B \right]^* = - (T^I)\du B
A$, {\it etc.}

The local transformation rule for the SM-I is generalized
from the global case (2.3) as
$$\eqalign{\d\phi^\a = \,& + 2(\e^A \psi\ud a A) V\du a\a~~, ~~~~
\d\Bar\phi^{\Bar\a} = +2 (\Bar\e^A \Bar\psi_{a A}) V^{a\Bar\a} ~~, \cr
\d\varphi^a = \,& + 2i ( \Bar\e^A \psi\ud a A) +(\d_Q\phi^{\ul\a})
A\du{\ul\a b} a \varphi^b - i (\d_Q\phi^{\ul\a}) A_{\ul\a} \varphi^a ~~, \cr
\d\varphi^*_a = \,& +2i (\e^A \Bar\psi_{a A}) - (\d_Q\phi^{\ul\a})
A\du{\ul\a a} b \varphi^*_b + i (\d_Q\phi^{\ul\a}) A_{\ul\a} \varphi^*_a ~~,
\cr
\d \psi^{a A} = \,&  - i \g^\m \Bar\e^A V\du \a a \Hat D_\m \phi^\a + \g^\m
\e^A
  \HatD_\m \varphi^a
  + (\d_Q \phi^{\ul\a}) A\du{\ul\a b}a  \psi^{b A} - i (\d_Q \phi^{\ul\a})
  A_{\ul\a} \psi^{a A} \cr
  & + \half\L_{\rm M} \psi^{a A} + \half \L_{\rm D} \psi^{a A} - \L^I
  (T^I\psi_a) _A ~~, \cr
\d\Bar\psi_{a A} = \,& + i \g^\m \e_A V_{\Bar\a a} \Hat D_\m
  \Bar\phi{\,}^{\Bar\a} -
  \g^\m\Bar\e_A\HatD_\m \varphi^*_a - (\d_Q \phi^{\ul\a}) A\du{\ul\a a} b
  \Bar\psi_{b A} + i (\d_Q \phi^{\ul\a})
  A_{\ul\a} \Bar\psi_{a A} \cr
  & + \half\L_{\rm M} \Bar\psi_{a A} + \half \L_{\rm D} \Bar\psi_{a A} - \L^I
  (T^I\, \Bar\psi_a) _A ~~. \cr }
\eqno(3.9) $$
As usual, all the covariant derivatives with {\it hats} are
$~Q\-$supercovariant
derivatives, {\it e.g.},
$$\Hat D_\m \varphi^a \equiv \calD_\m \varphi^a - 2i (\Bar\psi\du\m
A\psi\ud a A) \equiv \partial_\m\varphi^a - (\partial_\m\phi^{\ul\a})
A\du{\ul\a b} a \varphi^b + i(\partial_\m\phi^{\ul\a} ) A_{\ul\a}
\varphi^a - 2i(\Bar\psi\du\m A\psi\ud a A) ~~,
\eqno(3.10)$$
where $~\calD_\m\varphi^a$~ is the target space-time covariant
derivative like (3.6).  The only subtlety to be mentioned is the composite
connection such as $~(\partial_\m\phi^{\ul\a}) A\du{\ul\a b} a \phi^b$.
In the above supercovariantization prescription, the first
factor $~(\partial_\m\phi^{\ul\a})$~
should {\it not} to be supercovariantized, and this is independent of the
statistics of the field that the composite connection is acting on.  For
more detailed explanation related to quaternionic terms, see
{\it e.g.}~refs.~\bw\ref\ns{H.~Nishino and E.~Sezgin,
\np{278}{96}{353}.}.

We are now ready to present the invariant lagrangian of the
heterotic $~\s\-$model for the SM-I coupled to the $~N=(4,0)$~ superconformal
gravity:
$$\li{ e^{-1} \Lag_{\rm I} = \, & + g^{\m\n}
g_{\a\Bar\b}(\partial_\m\phi^\a) (\partial_\n\Bar\phi{\,}^{\Bar\b}) + g^{\m\n}
(\calD_\m\varphi^a) (\calD_\n\varphi^*_a)
+ \e^{\m\n} g_{\a\Bar\b} (\partial_\m\phi^\a)
(\partial_\n\Bar\phi{\,}^{\Bar\b})
\cr
& + i \left(\psi^{a A} \g^\m\calD_\m\Bar\psi_{a A}\right) + i \left(\Bar\psi_{a
A} \g^\m \calD_\m \psi^{a A} \right) \cr
& - \left(\Bar\psi\du\m A \g^\n\g^\m\Bar\psi_{a A}\right) V\du\a a
\left(\partial_\n\phi^\a + \HatD_\n\phi^\a \right)
- \left(\psi\du\m A \g^\n\g^\m\psi\ud a A\right) V_{\Bar\a a}
\left(\partial_\n\Bar\phi{\,}^{\Bar\a} + \HatD_\n\Bar\phi{\,}^{\Bar\a}
\right) \cr
& - i \left(\psi\du\m A \g^\n\g^\m\Bar\psi_{a A}\right)
\left(\calD_\n\varphi^a + \HatD_\n\varphi^a \right)
- i \left(\Bar\psi\du\m A \g^\n\g^\m\psi\ud a A\right)
\left(\calD_\n\varphi^*_a + \HatD_\n\varphi^*_a \right) {~~}.{~~~}
(3.11)  \cr} $$
The covariant derivative for $~\psi^{a A}$~ is defined by
$$\eqalign{{\cal D}_\m \psi^{a A} \equiv &\, \partial_\m \psi^{a A} - \half
\omega_\m \psi^{a A} + B\du\m I (T^I \psi^a)^A
- (\partial_\m \phi^{\ul\a}) A\du{\ul\a b} a \psi^{b A}
+ i (\partial_\m \phi^{\ul\a}) A_{\ul\a} \psi^{a A} ~~.  \cr }
\eqno(3.12) $$

The superinvariance of the action, including the quartic fermionic terms, can
be
easily confirmed by inspecting the supercovariance of all the fermionic field
equations.  It is worthwhile to mention that the
the gravitino field equation directly obtained from
the lagrangian (3.11) generates the gradient term $~\partial_\m\e$~ under
the $~Q\-$supertranslation, indicating its {\it non}-supercovariance.
However, this is superficial and poses no problem, because it turns
out to be proportional to the $~B\du\m I\-$field equation which is nothing but
the vanishing $~SU(2)$~ current itself.

The absence of purely matter fermionic quartic terms in the lagrangian is a
natural feature of our $~N=(4,0)$~ heterotic $~\s\-$model.  This is
because all the matter fermions $~\r^{-a}$~ or $~\pi^{-a}$~ carry the same
negative chirality, considering also the identity
$~(\g_m)\low{-\,-}(\g^m)\low{-\,-}\equiv 0$, it is impossible to form a Lorentz
invariant term out of purely matter fermions.

Note that the heterotic $~\s\-$model we have presented here shares the
same feature with $~N=2$~ heterotic $~\s\-$models \ghr\ {\it via} the
structure of the torsion tensor (3.3) or (3.4) as well as with other
$~N=(4,0)$~ heterotic $~\s\-$models with quaternionic K\"ahler structure \bw.
The limit to the flat target space-time is smooth by making the $~SU(n)$~
curvature vanish, because there is no restriction on the curvature tensor,
unlike the quaternionic K\"ahler manifold in the conventional
$~N=4\,~\s\-$models \pn\bs.

As usual in heterotic systems \bs\bns, we can add some Majorana-Weyl
unidexterous fermion (UF) denoted by $~\chi^{(r)}$~ with the positive chirality
to the $~\s\-$model system above.  Its invariant lagrangian is
$$\eqalign{e^{-1} \Lag_{\rm U F} = & \, + \fracm i 2 \left(\chi^{(r)}
\g^\m {\cal D}_\m \chi ^{(r)}\right) + \fracm 14
\left( \chi^{(r)} \g^\m \chi^{(s)} \right) \left[\,
2\left(\psi^{a A}\g_\m \Bar\psi\low {b A}\right) +
i\left(\varphi^a\partial_\m\varphi^*_b - \varphi^*_b\partial_\m\varphi^a\right)
\,\right] F\du a {b\,(r)(s)} \cr
& \, - \fracm i4 \left( \chi^{(r)} \g^m \chi^{(s)} \right) \left[\,
(\psi^\a \g_m \psi^\b ) \varphi^{*\Bar\g} \calD_\a F_{\b\Bar\g} +
(\Bar\psi{\,}^{\Bar\a} \g_m \Bar\psi{}^{\Bar\b}) \varphi^\g \calD_{\Bar\a}
F_{\g\Bar\b} \,\right] ~~, \cr }
\eqno(3.13) $$
up to sixth-order terms in fundamental fields.
The indices $~{\scst (r),~(s),~\cdots}$~ are for any arbitrary
representation of the target space-time Yang-Mills group, distinguished  from
the local Lorentz indices $~{\scst m,~n,~\cdots}$, and
$~\psi^\a\equiv\psi^a V\du a\a,~\varphi^\a\equiv
\varphi^a V\du a\a$.  The covariant derivative
for $~\chi^{(r)}$~ is
$${\cal D}_\m \chi^{(r)} \equiv \partial_\m\psi^{(r)} + \fracm1 2 \omega_\m
\chi^{(r)} + (\partial_\m \phi^\a) A_{\a}{}^{(r)(s)} \chi^{(s)} +
(\partial_\m \Bar\phi{\,}^{\Bar\a}) A_{\Bar\a}{}^{(r)(s)} \chi^{(s)} ~~.
\eqno(3.14) $$
The $~A\du{\ul\a}{(r)(s)}$~ are the
Yang-Mills composite gauge field  in the target space-time whose field strength
is $~F\du{\ul\a\ul\b}{\,(r)(s)}$, while $~F\du a {b\,(r)(s)}\equiv
F_{\a\Bar\b}{\,}^{(r)(s)} V\du a \a V^{b\Bar\b}$.
The transformation rule for $~\chi^{(r)}$~ is
$$\d\chi^{(r)} \equiv -(\d_Q\phi^{\ul\a}) A\du{\ul\a}{(r)(s)}\chi^{(s)}
 + \half\left[ (\d_Q \varphi^a) \varphi^*_b - (\d_Q \varphi^*_b) \varphi^a
\right] F\du a{b\,(r)(s)} \chi^{(s)}
+\half \L_{\rm D} \chi^{(r)} - \half \L_{\rm M} \chi^{(r)}~, ~~
(3.15) $$
up to quartic order terms in fundamental fields.

As usual \ref\hw{C.~Hull and E.~Witten, \pl{150}{85}{398}.},
the superinvariance of $~\Lag_{\rm UF}$~ requires that
$$~F_{\a\b}{}^{(r)(s)}= F_{\Bar\a\Bar\b}{}^{(r)(s)}= 0~~.
\eqno(3.16) $$
We have confirmed the superinvariance of the action (3.13) up to the
sixth-order terms in the fundamental fields multiplied by
$~F_{\a\Bar\b}{\,}^{(r)(s)}$~ or
its covariant derivatives.  In the invariance check, it is convenient to
go to a special frame where all the composite connection (gauge) fields
vanish, except for their curvatures (field strengths).  For example, by adding
a  composite gauge transformation to (3.15), we can delete completely the
r.h.s.~of $~\d_Q\chi^{(r)}$.  We use this convenient frame which is
analogous to the ``geodesic frame'' used in general relativity, in order to
simplify drastically all the computations.  In this frame, all the
transformation rules (3.9) and (3.15) contain only the composite curvatures
with {\it no} bare  composite connection fields, and therefore all the terms
with these bare composite fields after the variation become unimportant.
Since $~\varphi^a$~ is dimensionless,
its higher-order terms may arise at any order.  However, we are confident about
the perturbative superinvariance at higher orders in the basic fields.

Even though the heterotic $~\s\-$model $~\Lag_{\rm I}$~ has a structure
different from the already-known theories, the UF piece (3.13) has
the same feature as the known ones \bs\ as well as differences.  In particular,
the target space-time Yang-Mills fields has the same couplings to the
UF, while the bosonic terms $~\varphi\partial\varphi^*
- \varphi^*\partial\varphi$~ also enters into this coupling accompanying the
usual $~(\psi\g\Bar\psi)$~ to form the $~SU(n)$~ current
combination.  Relevantly, the term $~\left[ (\d_Q\varphi) \varphi^* -
(\d_Q\varphi^*)\varphi \right]F\chi$~ is needed in (3.15).  In fact, it is
this term that guarantees the supercovariance of the $~\chi\-$field
equation, which would have generated the gradient term $~\partial_\m\e$~ in its
$~Q\-$supertranslation, if this extra term were absent in $~\d\low Q\chi$.

\bigskip\bigskip\bigskip


\centerline{\bf 4.~~Concluding Remarks}

Based on the new matter scalar multiplets obtained out of $~D=1$~ by the
dimensional oxidation technique \gr, we have presented an alternative $~N=4$~
superstring $~\s\-$model with field contents different from any
previously-known
$~N=4$~ superstring theories.   We have established convenient categorization
of
possible scalar multiplets which may couple to superconformal gravity.  There
are four possible $~N=4$~ matter scalar multiplets, and two of them are
promoted
to the previously-known $~\s\-$models in  refs.~\pn\bs, when coupled to
superconformal gravity.

We have seen that our K\"ahler
$~\s\-$model based on the \hbox{SM-I} has a mixed property of $~N=(2,0)$~
heterotic  $~\s\-$models \ghr\bns\ and other previous $~N=(4,0)$~ heterotic
$~\s\-$models  \pn\bs.  The possibility of the non-vanishing torsion tensor,
which is proportional to the curl of the antisymmetric tensor in the target
space-time, is like the $~N=(2,0)$~ \ghr\bns\ or $~N=(4,0)$~ \bs\ heterotic
$~\s\-$models.

Another interesting aspects we found in our investigation is that
when we couple global matter multiplets to superconformal gravity, the presence
of minimal couplings between the $~SU(2)$~ gauge field in the supergravity
multiplet and matter scalars in non-trivial $~SU(2)$~ representations seems to
cause troubles for superconformal invariance.  In terms of the four scalar
matter multiplets SM-I through SM-IV, we have seen that the SM-II and SM-IV
correspond to the heterotic $~\s\-$model in ref.~\bs, while the SM-I is
promoted to our new heterotic $~\s\-$model on a K\"ahler manifold.
Since all the scalars in the SM-III are non-singlets under the $~SU(2)$, this
multiplet appears to have the problem with coupling to
superconformal gravity.

We stress the importance of our results even for the superconformal
gravity (3.1) itself, with the gravitini in the simple complex ${\bf 2}
\-$representation of the $~SU(2)$~ group, which has not been presented
before.\footnotew{For example in \pn\ the gravitini carry the same
$~{\bf 2}$-indices, but they also have additional $~Sp(1)$~ indices subject
to what is called symplectic Majorana condition.}  This particular
representation  was motivated by the matter multiplets SM-I through SM-IV
obtained by  the dimensional oxidation.  Furthermore, the scalar
representation in SM-I motivates the study of the complex coset
K\"ahler manifold $~SU(n,1) / SU(n)
\otimes U(1)$~ as the appropriate $~\s\-$model
among other potentially possible K\"ahler manifolds for $~N=(4,0)$~ superstring
models.

It is also to be emphasized that one of the two scalars in the
global SM-I multiplet (2.1) is promoted to be the coordinates of this coset,
while the other one plays a role of a tangent vector under the holonomy group
$~SU(n)\otimes U(1)$.  This is a new peculiar feature discovered in the study
of our $~N=(4,0)$~ heterotic $~\s\-$model, which has not been presented in the
past to our knowledge.  The explicit technical treatment of complex coordinates
for the  K\"ahler manifold given in this paper has clarified the subtlety
related to complex-conjugations, which will be also useful in the future.
Another interesting feature of our model is the peculiar UF lagrangian (3.13),
where the $~SU(n)$~  current couples to the target space-time Yang-Mills field
strength.

Even though we chose in this paper the representation $~\psi^{a A}$~ for
the fermion, it is to be straightforward to rewrite the whole
system in terms of $~\psi^{\a A}$~ with the curved target space-time curved
index $~{\scst\a}$.  In such a case, the system will share more common
technical features with the model in ref.~\bs.

According to the conventional wisdom \pn\bs, only those hyper-K\"ahler
manifolds, that can be promoted to be quaternionic K\"ahler manifolds,
are supposed to be the suitable cosets for $~N=(4,0)$~ heterotic
$~\s\-$models.  In this paper, we have established a completely new
possibility of $~N=(4,0)$~ heterotic $~\s\-$models
that requires such simple cosets as $~SU(n,1) / SU(n) \otimes U(1)$.  This
coset
is just a simple example of other possible cosets, which are compatible
with our complex-representations in SM-I.  Especially, we found that the
dimensionality of the target space-time for the $~N=4$~ superstring has to be
no longer a multiple of four in real coordinates, but it can be an even
integer.  This coincides with the intuitive understanding that the number
of scalar fields parametrizing the target manifold in our $~N=(4,0)$~ system
is half of the conventional one \pn\bs, leaving the remaining super-partner
scalar fields living on its tangent space.
It is also very suggestive that the recently discovered discrete symmetry
$~SL(2,\ZZ) \subset SL(2,\IR)$~ called $~S\-$duality \ref\quevedo{A.~Font,
L.~Iba\~nez, D.~Lust and F.~Quevedo,  \pl{249}{90}{35};\\ S.J.~Rey,
\pr{43}{91}{526}.} is also related to our coset {\it via} $~SL(2,\IR) / SO(2)
\approx SU(1,1) / U(1)$, because $~N=4$~ superstring
theory may well be the fundamental underlying theory for ordinary $~N=1$~
superstring theory, just like the  $~N=2$~ case \kng.

The purpose of investigating heterotic $~\s\-$models coupled to superconformal
gravity is to seek the consistent realization of $~N=4$~ superconformal
algebra with explicit representations of matter multiplets.  It is these
explicit representations that enable us to promote the superconformal
algebra to a string theory.  In fact, even though superconformal algebra
exists for {\it arbitrary} $~N$~ up to $~N=\infty$~ \bns,
there is such restriction as $~N\le 4$~ for realizing superstring theories
due to the consistency of the matter representation coupled to superconformal
gravity.  It  seems that a superconformal algebra {\it without} matter
representations is {\it not} explicit enough to construct a string theory.

As was already mentioned, the possible critical dimension $~D_c=+4$~
\siegelcritical\ provides the strong motivation for the intensive study of the
$~N=4$~ superstring, not only for usual applications to high-energy particle
physics, but also as an important superconformal theory for lower-dimensional
supersymmetric integrable systems \kng.

\bigskip\bigskip\bigskip

We are grateful to S.J.~Gates, Jr.~for his providing the basic idea,
valuable discussions, useful communications, helpful suggestions,
important remarks, and careful reading of the manuscript.

\bigskip\bigskip

\vfill\eject

\footatend\vfill\supereject\immediate\closeout\rfile\writestoppt
\baselineskip=14pt\centerline{{\bf References}}\bigskip{\frenchspacing%
\parindent=20pt\escapechar=` \input refs.tmp\vfill\eject}\nonfrenchspacing

\end{document}